\documentclass{jaa}

\usepackage{graphicx}
\usepackage{color}

\begin{document}

\title{Searching for Dual Active Galactic Nuclei}


\author{Rubinur K.\textsuperscript{1,2}, M. Das\textsuperscript{1} \and P. Kharb\textsuperscript{3}}
\affilOne{\textsuperscript{1}Indian Institute of Astrophysics, Bangalore 560034, India.\\}
\affilTwo{\textsuperscript{2}Pondicherry University, R. Venkataraman Nagar, Kalapet, 605014 Pondicherry, India.\\}
\affilThree{\textsuperscript{3}National Centre for Radio Astrophysics - Tata Institute of Fundamental Research, S. P. Pune University Campus, Post Bag 3, Ganeshkhind Pune 411007, India.}


\twocolumn[{

\maketitle

\corres{rubinur@iiap.res.in}


\begin{abstract}
Binary or dual active galactic nuclei (DAGN) are expected from galaxy formation theories. 
However, confirmed DAGN are rare and finding these systems has proved to be challenging. 
Recent systematic searches for DAGN using double-peaked emission lines have yielded several new detections. 
As have the studies of samples of merging galaxies. In this paper, 
we present an updated list of DAGN compiled from published data. 
We also present preliminary results from our ongoing Expanded Very Large Array (EVLA) radio study
of eight double-peaked emission-line AGN (DPAGN). One of the sample galaxy shows an S-shaped radio jet. 
Using new and archival data, we have successfully fitted a precessing jet model to this radio source. 
We find that the jet precession could be due to a binary AGN with a super-massive black-hole (SMBH) 
separation of $\sim$0.02~pc or a single AGN with a tilted accretion disk. 
We have found that another sample galaxy, which is undergoing a merger, 
has two radio cores with a projected separation of 5.6 kpc. 
We discuss the preliminary results from our radio study.
\end{abstract}

 }]


\doinum{12.3456/s78910-011-012-3}
\artcitid{\#\#\#\#}
\volnum{123}
\year{2016}
\pgrange{23--25}
\setcounter{page}{23}
\lp{25}

\section{Introduction}
According to the ${\Lambda}\mathrm{CDM}$ cosmological model, massive galaxies grow through
the mergers of smaller galaxies (Springel et al. 2005). We also now believe that almost
all galaxies have a super-massive black-hole (SMBH) at their center. 
Therefore when two galaxies merge, their SMBHs fall into the center of the merger remnant. 
Simulations show that galaxy mergers cause the gas to flow towards the center (Hopkins \& Hernquist 2009).
This gas inflow can ignite the accretion activity of the SMBHs and turn them into active galactic nuclei (AGN).
Hence, we expect AGN pairs at the centers of merger remnants. 
These AGN pairs are known as dual AGN (DAGN) when the separation is 100~parsec to 10~kpc and binary AGN 
when the separation is $<$100~parsec (Burke-Spolaor et al. 2014). 
While we expect a large number of dual or binary AGN, there are very few detections till date.\\

Dual or binary AGN can be detected using high resolution optical, IR, radio or X-ray imaging and spectroscopy. 
The radio source 3C~75 (Hudson et al. 2006) and the ultra-luminous infrared galaxy NGC~6240 (Komossa et al. 2003)
are examples of well-known DAGN. So far, there are only two known binary AGN candidates detected through
the technique of Very Long Baseline Interferometry (VLBI): 4C~+37.11 (Rodriguez et al. 2006) 
and NGC~7674 (Kharb et al. 2017a). However, these discoveries have been  serendipitous. 
After the release of Sloan Digital Sky Survey (SDSS\footnote{www.sdss.org}) spectroscopic data,
large numbers of active galaxies with double-peaked narrow emission lines have been discovered. 
These are known as double-peaked AGN (DPAGN, Zhou et al. 2004). DPAGN can arise from dual AGN 
as well as from jet-medium interaction (Rosario et al. 2010; Kharb et al. 2015; Kharb et al. 2017b) 
or rotating disks (Greene \& Ho 2005). High-resolution imaging needs to be carried out to confirm their DAGN nature.\\

There are other indirect signatures such as S- or X shaped radio sources (Begelman, Blandford \& Rees 1980)
or periodicity in optical variability (e.g. OJ287 Sillanpaa et al. 1988; Lehto \& Valtonen 1996). 
 While detecting dual/binary AGN is extremely challenging, these systems are important to 
understand galaxy evolution during the final stages of the galaxy merging process.\\

While two radio nuclei are sometimes detected in merging or merged galaxies, 
they can be any of the following: (i) binary/dual AGN, (ii) a single AGN + a single star-forming (SF) nuclei
or (iii) dual SF nuclei. To confirm the presence of two AGN, high-resolution radio or X-ray observations are required. 
At radio frequencies, the two nuclei should be compact and have flat or inverted spectral indices (Hovatta et al 2014).\\

In this paper, we present a list of confirmed DAGN or binary AGN compiled from the literature.
We also discuss results from the multi-frequency Expanded Very Large Array (EVLA) radio study of our DPAGN sample.
We discuss preliminary results from this study.

\section{Dual AGN in the Literature}
\subsection{Sample Selection Methods}
Most of the initial detections of dual AGN were serendipitous, and involved direct imaging in different wavelengths
. However, Zhou et al. (2004) detected a dual AGN candidate in an SDSS double-peaked emission line galaxy.
Later Gerke et al. (2007) and Comerford et al. (2009a) started the systematic search of dual/binary AGN candidates
using the spatially resolved double-peaked [O III] emission line in DEEP2 galaxy redshift survey\footnote{deep.ps.uci.edu/}
(Davis et al. 2004). Around the same time, SDSS DR7 data was released and large number of samples 
from double-peaked emission line galaxies were assembled (Wang et al. 2009, Smith et al. 2009, Ge et al. 2012, Fu et al. 2012)
. These authors examined different properties of double-peaked emission lines to obtain the best dual AGN candidates.\\ 

Ge et al. (2012) made an automated pipeline to select DPAGN.
They created a sample of 3030 DPAGN using both the pipeline and visual inspection. Wang et al. (2009) assembled
a sample of 87 type 2 DPAGN and found a correlation between the ratios of the blueshift to the redshift 
of the emission line ([O III]) and the double peak line fluxes. 
This correlation could be explained by the Keplerian relation predicted by models of co-rotating dual AGN.
Therefore, they claimed that most of their DPAGN were good dual AGN candidates 
and not rotating disks or outflows. Fu et al. (2012) and  Fu et al. (2011a) observed 42 DPAGN with integral field spectroscopy
and high-resolution optical and NIR imaging (sub-arcsec). 
Comerford et al. (2012) included long slit spectroscopy of DPAGN to select DAGN candidates.\\

There are samples of DAGN candidates that have been derived from merging galaxies as observed in optical images.
Koss et al. (2012) used the Swift Burst Alert Telescope (BAT\footnote{http://swift.gsfc.nasa.gov/archive})
data as well as optical data to study merging systems. 
Teng et al. (2012) made a pilot study with the Chandra X-ray Observatory \footnote{chandra.harvard.edu/}
of 12 massive galaxy mergers. Mezcua et al. (2014) studied a sample of 52 disk galaxies 
which are considered to be candidates for minor merger events and 
created a sample of 19 sub-kpc separation dual AGN candidates. 
Satyapal et al. (2017) used the Wide-field Infrared Survey Explorer (WISE: Wright et al. 2010)
infrared color to select the DAGN candidates. According to them, 
the sample selection using optical spectroscopy may not be a good idea as there are many obscured AGN 
which display optical spectra similar to star-forming nuclei.

\subsection{The Detection of Dual AGN}
Double peaked emission line galaxies appeared to be good candidates for detecting DAGN (Zhou et al. 2004).
The search for dual AGN in DPAGN has been going on for more than a decade. 
The confirmation finally comes from high resolution (sub-arcsec to milli-arcsec) imaging of DAGN candidates. 
Sources showing S-shaped radio jets or periodicities in their optical curves are likely
to have very close binary black holes, which may or may not be resolved by existing telescopes
(Rubinur et al. 2017 and reference therein). \\

Following are examples of DAGN detections from studies of DPAGN. Liu et al. (2010)
discovered four dual AGN using high-resolution imaging of double-peaked emission line candidates.
Fu et al. (2012) discovered one dual AGN. Comerford et al. (2011) detected a dual AGN using
Chandra observations of a double-peaked emission line galaxy. M\"{u}ller-S\'{a}nchez et al. (2015)
detected four DAGN. McGurk et al. (2015) detected one DAGN and noted that 0.05\% of the SDSS
DPAGN are spectroscopic dual AGNs within 3$^{\prime\prime}$. 
However, there is also a growing consensus that DPAGN studies are not the most efficient way 
to discover DAGN (Comerford et al. 2012; Fu et al. 2012; M\"{u}ller-S\'{a}nchez et al. 2015).\\ 

On the other hand, the sample selection from optical imaging of merging systems has yielded several DAGN detections. 
Koss et al. (2012) have detected 5 DAGN using SWIFT BAT data and optical merger systems. 
Fu et al. (2015) selected DAGN candidates from radio imaging and confirmed 4 DAGN from 6 candidates. 
Satyapal et al. (2017) have selected the sample from mid-IR color using WISE observations where they have detected 4 DAGN. 
The recently detected DAGN, SDSS J140737.17+442856.2, is also from the same method (Ellison et al. 2017).\\

Das et al. (2017) have presented a list of DAGN where they have noted 26 confirmed DAGN.
However, that list is incomplete and does not include some recent detections. 
We present an updated list of confirmed DAGN in Table~1.

\section{Our Sample Selection and Observations}
We have selected our target galaxies from the sample of 3030 DPAGN in Ge et al. (2012) and Fu et al. (2012). 
The sample comprised of only narrow line DPAGN sources.
We used the velocity difference of the red-shifted and blue-shifted [O III] lines ($\Delta V$) and
the stellar velocity dispersion ($\sigma_\star$) to select the sample. 
We found that most of the confirmed DAGN had $\Delta V$ $\geq$ 400 km~s$^{-1}$ 
in the  $\Delta V$ vs. $\sigma_\star$ correlation. Therefore, we chose sources which were 
(1) offset from the $\sigma_\star$ $\propto \Delta$ V correlation, 
(2) satisfied the condition $\Delta V$ $\geq$ 400 km~s$^{-1}$, 
(3) had redshifts z~$\leq$~0.1, (4) had disky morphology in SDSS optical images, 
and (5) had radio emission in FIRST or NVSS images (Rubinur et al. 2018, in preparation).
 We found 6 DPAGN from Ge et al. (2012) which satisfied the above criteria. We chose 2 DPAGN from Fu et al. (2012) which also satisfied the above criteria but one of the objects had $\Delta V = 307$~km~s$^{-1}$.

We observed 8 DPAGN with the EVLA at 6 GHz (C-band) in the A-array configuration
for 1.5 hours on 19th July, 2015 (VLA/15A-068). 
The observations were carried out with 1792 MHz wide baseband centered at 5.935 GHz
with fourteen spectral windows, each window had 64 channels with a frequency resolution of 2 MHz.
Each target was observed for $\sim$4 minutes, with 50~s scans on the phase calibrators.

Based on the results of the 6~GHz observations, we observed 5 of the 8 galaxies (Table~2)
with the EVLA at 15 GHz (U-band) in the B-array configuration in 2016 (VLA/16A-144). 
We also analyzed 8-12 GHz (X-band; 13B-020, PI: J. M. Comerford) archival data for two of our sample sources.
We used the Common Astronomy Software Applications (CASA\footnote{https://casa.nrao.edu/}) (McMullin et al. 2007) 
and Astronomical Image Processing System (AIPS\footnote{www.aips.nrao.edu/}) packages
for the EVLA data-reduction and analysis (see details in Rubinur et al 2017).


\section{Results from Our Observations}
We have now completed the analysis of 6~GHz data of 8 DPAGN. 
We find that two galaxies have compact single cores, four galaxies have single but elongated radio cores, 
which could suggest the presence of unresolved jets in them and the remaining two sources show 
dual radio components. We carried out follow-up observations for the elongated and 
dual component sources. We obtained 15~GHz observations for three elongated sources. 
The fourth one already had VLA X-band (8-12~GHz) archival data whose resolution was comparable 
to the 15~GHz observations. For the two dual component sources we obtained 15~GHz data and 
for one of them (2MASX J12032061+1319316), we found archival X (8-12) band data as well.


The $6-15$ GHz spectral index maps suggest that two of the three elongated galaxies have
radio jets but they are not resolved at 6~GHz or 15~GHz. 
The third source has a flat spectral index consistent with a compact AGN. 
The $8-12$ GHz spectral index of the elongated source is steep which imply
that it is a core-jet structure. Below we discuss the two galaxies with dual radio components.\\

\begin{figure}[!t]
\centerline{\includegraphics[width=10.2cm]{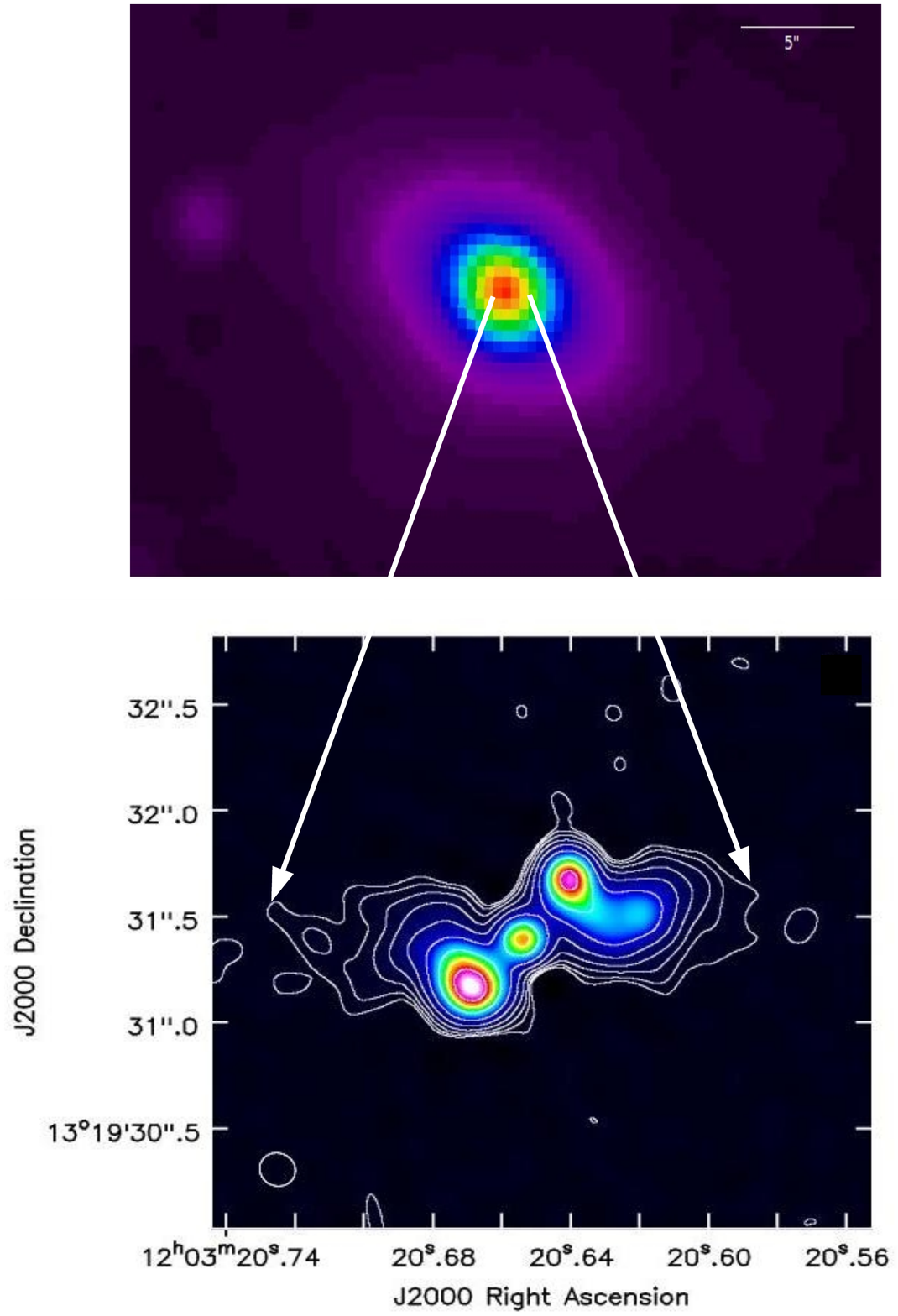}}
\caption{The top panel shows the SDSS DR13 (Albareti et al. 2016) r-band optical image of the DPAGN,
2MASX~J12032061+13193.
This DPAGN has an S-shaped radio jet which is well-fitted by the precessing jet model. 
The jet has a total extent of $\sim$3 kpc. The bottom panel shows the zoomed-in version 
of the radio color image with radio contours. This is a total intensity map at 8-12~GHz. 
The beam-size is $0.16^{\prime\prime}\times0.15^{\prime\prime}$. 
The rms noise in the image is $\approx 10~\mu$Jy~beam$^{-1}$. 
The contour levels correspond to 0.60, 1.25, 2.5, 5, 10, 20, 40, 60, 80\% of the 
peak flux density value of 5.2 mJy~beam$^{-1}$. For details see Rubinur et al. (2017).}\label{figOne}
\end{figure}

\subsection{The Galaxy with an S-shaped Radio Jet}
The DPAGN, 2MASX~J12032061+1319316 (J1203 for short), showed two compact radio components
in its 6~GHz EVLA image (Rubinur et al. 2017). The components were separated by 0.7$^{\prime\prime}$.
We obtained higher-resolution 8-12~GHz EVLA archival data of this galaxy and found that the two components
were actually two hot-spots of an S-shaped radio jet (see Figure 1). 
The core size is $\sim$ 100 parsec and the total extent of the jets is $\sim$3 kpc. 
As the 15 GHz observations were obtained in the B-array configuration of the EVLA, 
these data had a resolution similar to the 6 GHz data; images showed the two hot-spots but no clear radio core in between.\\

\subsection{The Dual Core Galaxy}
The optical SDSS image of J1617+2226 shows that one of the sample DPAGN is
undergoing a galaxy merger with a smaller companion. Its 6~GHz EVLA image shows that this DPAGN
has two radio cores associated with two optical cores. 
The projected separation between the radio cores is $\sim$5.6~kpc. 
The 15~GHz EVLA image also shows the two radio cores. 
The primary core (associated with the DPAGN) has a small one-sided jet which is not clearly detected
in the 6~GHz image. We have used both the images to make the spectral index image in order to understand
the emission mechanism in both the cores. The average spectral index ($\alpha$, 
defined such that flux density at frequency $\nu$ is S${_\nu}$ $\propto \nu^{\alpha}$) 
of the primary core is $-0.8$ and the secondary core is $-0.3$.\\ 


\section{Discussion}
We have found that one of the sources in our sample shows an S-shaped core-jet structure and another source shows dual radio cores in radio images. Here we discuss possible mechanisms that can explain these radio structures.
\subsection{ J1203: S-shaped core-jet structure} We have fitted the helical jet model of Hjellming \& Johnston (1981) to the observed S-shaped radio structure. This model does not assume any particular origin of the jet precession. We have obtained a precession period of 10$^5$ yrs for the jet  (see details in Rubinur et al. 2017). We have calculated the age of the jet using a simplistic spectral aging model. We have calculated the total radio luminosity (O’Dea  \& Owen 1987) and used that to calculate the minimum-energy magnetic field (B$_{min}$; Burbidge 1959) which is close to equipartition of the energy in particles and magnetic fields. We estimated the magnetic field of 105~$\mu$G at 11.5~GHz. We have used the relation from van der Laan \& Perola (1969) to derive the electron lifetimes; these also turn out to be $\sim$10$^5$ yrs. As the precession and spectral aging models give the same age of $\sim$10$^5$ yrs, we concluded that J1203 is a young radio source and most likely a compact symmetric object (CSO; see details in Rubinur et al. 2017). CSOs are radio sources with compact symmetric double lobes that extend to $\leq$ 1 kpc (Wilkinson et al.1994).\\

Jet precession can arise due to (i) the presence of binary supermassive black holes. We have used the relations in Begelman, Blandford \& Rees (1980) to calculate the separation of the binary SMBH (if present) using the precession period obtained from the fitted model and calculated the SMBH mass from the M~$-~\sigma_{\star}$ relation. This calculation suggests a SMBH separation of 0.02~parsec, which is 18 micro-arcseconds in the sky. Such separations are not resolvable with the present day telescopes. (ii) Lu (1990) have suggested that tilted accretion disks of single SMBHs can also give rise to jet precession. We used the M$_\mathrm{abs}~-$~P relation from Lu (1990) where M$_\mathrm{abs}$ is the absolute magnitude in the B band (445 nm) and P is the precession period. We used the observed magnitude from NED\footnote{https://ned.ipac.caltech.edu//} and calculated the period to be between $10^{5}$ and $10^{9}$ yrs. Therefore, there was a large uncertainty in the calculated period, even though the estimated period lay in the same range. (iii) The third possibility for a precessing jet could be that there is a dual SMBH system which is well separated, but the secondary SMBH, which is not detected in the radio, has passed by the primary SMBH in the past. This could also introduce the precession in the jet of the primary SMBH (Rubinur et al. 2017).\\

\subsection{J1617: Dual AGN or not} We propose that the steep spectral index of the primary core is because of the presence of an unresolved jet. The secondary core with the flatter spectral index could either be due to a compact AGN emitting synchrotron radiation, or thermal free-free emission from a star-forming nuclei. Both the cores have SDSS optical spectra. The spectrum of the main core indeed resembles an AGN, while the spectrum of the second core is consistent with star-formation (Rubinur et al. 2018, in preparation).\\

\subsection{Why is the detection of binary/dual AGN important?}
The primary motivation to detect DAGN and binary AGN is to understand galaxy evolution 
and the final stages of galaxy mergers. These phenomena are still not well understood. 
The primary issue is that there are still not enough confirmed DAGN, although that number is steadily increasing.
Another issue is that DAGN are detected at different specific wavelengths 
but multi-wavelength data is required to understand the different processes associated with DAGN (like AGN feedback, star formation, gas/stellar disks). There are only a few DAGN that have multi-wavelength detections. The ultraluminous infrared galaxy (ULIRG) NGC~6240 is one such example; it is a gas rich merger remnant, with sensitive multi-wavelength observations (Wizinwich et al. 2000a, Max et al. 2007). These authors have found that there are two rotating stellar disks surrounding the two SMBHs.\\

Comerford et al. (2011) found that the DAGN SDSS J171544.05+600835.7 may have a bipolar outflow or a counter-rotating central disk which can cause the misalignment of gas and affect stellar kinematics. The host galaxies of DAGN are ideal laboratories to study the complex gas and stellar dynamics associated with galaxy  mergers. Recently, Shangguan et al. (2016) have carried out a deep study of 4 DAGN using the Hubble Space Telescope (HST\footnote{ nasa.gov/hubble}) and Chandra space telescopes. They found that three galaxies are undergoing major mergers while the remaining one has gone through a minor merger. They have estimated the masses of the SMBHs, bulge masses and Eddington ratios, and tried to compare their sample with a control sample that have single AGN. Interestingly, they did not find much difference between the DAGN sample and control sample with respect to the parameters listed above. \\

In this study we have focused on dual nuclei where both nuclei are  AGN. However, there are many merger remnants which have (AGN + SF) or (SF + SF) nuclei, as discussed in Section~1. In fact SF plays an important role in the evolution of dual nuclei systems and many questions remain regarding the role of stellar and AGN feedback in the merging process (Mazzarella et al 2012). How do galaxy mergers evolve with different types of nuclei? Does the SFR change with the presence of dual nuclei? How are the gas and stellar kinematics affected in the presence of dual nuclei? These questions can be addressed with sensitive, high-resolution optical or ultraviolet (UV) observations which can help us to understand star-formation in the presence of DAGN. We have recently started examining the UV emission from dual nuclei galaxies using the UV Imaging telescope (UVIT) on the ASTROSAT satellite (Rubinur et al. 2018, in preparation).

\section{Conclusions}
\begin{enumerate}
\item  We have complied a list of 36 confirmed dual and binary AGN from the literature. Systematic searches of DAGN were initiated using DPAGN samples, as well as galaxy merger systems as observed in optical images. It is found that the DAGN detection from the DPAGN samples is quite small. Searches of merging galaxies are giving better yields.
\item We have created a sample of DPAGN to detect DAGN. We have observed these galaxies using the EVLA at 6 and 15 GHz. The preliminary results of 8 galaxies show that two have compact single cores while four have elongated single cores, suggestive of unresolved core-jet structures. One source has an S-shaped jet structure (2MASX~J12032061+1319316) and one has dual radio nuclei.
\item We have closely examined the source J1203, which has an S-shaped radio jet structure. We have fitted the precession jet model and found a precession time-scale of 10$^5$ yrs. We find a similar time-scale from spectral aging. This source has a small jet of size 3 kpc and is a young compact symmetric object (CSO). The jet precession could be due to the presence of an unresolved binary AGN or a single AGN with a tilted accretion disk.
\item  The 6~GHz and 15~GHz EVLA observations of our dual core galaxy gives a hint of the presence of a DAGN with a separation of 5.6 kpc. A detailed multi-wavelength spectral analysis to understand the true nature of the dual cores is currently underway.
\item  A galaxy in the presence of dual nuclei can behave differently compared to that with a single AGN or AGN + SF nuclei. UV observations of merger galaxies can help us understand star formation in presence of different types of nuclei. We are currently analyzing these effects in our newly acquired UVIT ASTROSAT data.
\end{enumerate}

\section*{Acknowledgement}
We thank the referee for insightful comments that improved the
paper. The National Radio Astronomy Observatory is a facility of the National Science Foundation operated under cooperative agreement by Associated Universities, Inc. This research has made use of the NASA/IPAC Extragalactic Database (NED), which is operated by the Jet Propulsion Laboratory, California Institute of Technology, under contract with the National Aeronautics and Space Administration. Funding for the Sloan Digital Sky Survey IV has been provided by the Alfred P. Sloan Foundation, the U.S. Department of Energy Office of Science, and the Participating Institutions. SDSS- IV acknowledges support and resources from the Center for High-Performance Computing at the University of Utah. The
SDSS web site is www.sdss.org. We would like to thank the organizers of the RETCO-2017 meeting.



\onecolumn
\begin{table}
\tabularfont
\caption{List of confirmed binary and dual AGN compiled from the literature.}\label{tableExample} 
\begin{tabular}{|c|c|c|c|c|}
\topline
No & Name &Redshift & Projected Separation in kpc& Reference \\\hline
1& Was 49 & 0.060979 &  8.3 & Bothun et al. (1989)\\
2& LBQS 0103-2753 & 0.85 &  2.3 &Junkkarinen et al. (2001)\\
3& NGC~326 &0.047400 & 6.67  & Murgia et al. (2001)\\
4& NGC 6240 &0.024 &0.9 & Komossa et al. (2003)\\
5& 4C +37.11 & 0.055 & 0.007 & Rodriguez et al. (2006) \\
6& 3C~75  &0.023 &6.4  & Hudson et al. (2006)\\
7& MRK~463 &0.050 & 3.8 & Bianchi et al. (2008)\\
8& CID 42 &0.359 & 2.46 & Civano et al. (2010)\\
9& SDSS J150243.1+111557 & 0.39 & 7.4   & Fu et al. (2011)\\
10& SDSS J095207.62+25527.2 & 0.339 &  4.8   & McGurk et al. (2011) \\ 
11& SDSS J171544.02+600835.4 & 0.156300 & 1.9  & Comerford et al. (2011)\\
12&  SDSS J142607.71+353351.3 &1.175& 5.5 & Barrows et al. (2012)\\ 
13& MRK 739 & 0.02985 & 3.4 &  Koss et al. (2012)\\
14& IRAS 05589+2828 & 0.033000   & 8.0 & Koss et al. (2012)\\
15& ESO 509-IG066 NED 02 & 0.033223 & 10.5 & Koss et al. (2012)\\
16& IRAS 03219+4031  & -- & 10.8 & Koss et al. (2012)\\
17& NGC 3227 &  0.003859 & 12.3 & Koss et al. (2012)\\
18&  MRK 266 & 0.028 & 6.0 & Mazzarella et al. (2012) \\
19& SDSS J110713.22+650606.6 & 0.033 & 8.8 & Teng et al. (2012)\\
20& 2MASX J11085103+0659014 &0.1816  &2.1 & Liu et al. (2013)\\
21& SDSS J114642.47+511029.6  &  0.1300 & 6.3 & Liu et al. (2013)\\
22& 2MASX J00383316+4128509 &  0.073 &  4.7  & Huang et al. (2014)\\
23& SDSS J132323.33−015941.9 & 0.35   & 0.8   & Woo et al. (2014)\\
24& SDSS J112659.59+294442.8 &  0.101827  & 2.2 & Comerford et al. (2015)\\
25& SDSS J220635.08+000323.1 & 0.046100   &  4.6  & Fu et al. (2015)\\
26& SDSS J005113.92+002047.0 & 0.1126 & 3.4  & Fu et al. (2015)\\
27& SDSS J230010.16-000531.3 & 0.1797 &  2.5 & Fu et al. (2015)\\
28& SDSS J223222.41+001226.3 & 0.2210  & 3.2  & Fu et al. (2015)\\
29& SDSS J115822+323102 & 0.1658 & 0.62  & M\"{u}ller-S\'{a}nchez et al. (2015)\\
30& SDSS J162345+080851 &0.1992  &  1.55  & M\"{u}ller-S\'{a}nchez et al. (2015)\\
31& NGC~7674 & 0.028924 & 0.00035 & Kharb et al. (2017a)\\
32& SDSS J1407+4428  &0.143  & 8.3 & Ellison et al. (2017)\\
33& SDSS  J103631.88+022144.1 &0.05040 &  2.8 & Satyapal et al. (2017)\\
34& SDSS J130653.60+073518.1 & 0.11111 & 7.5   & Satyapal et al. (2017)\\
35& SDSSJ122104.98+113752.3 &  0.05546  & 9.3 & Satyapal et al. (2017)\\
36& SDSS 104518.03+351913.1  & 0.06758 &  9.0  & Satyapal et al. (2017)\\
\hline
\end{tabular}
\tablenotes{Note. Column 2 gives the name of the binary and dual AGN galaxies. Column 3 gives the redshift. 
Column 4 gives the projected separation of the dual/binary AGN in kpc. Column 5 notes the references corresponding to the DAGN detection.}
\end{table}

\begin{table}
\tabularfont
\caption{Observation details}\label{tableExample} 
\begin{tabular}{|c|c|c|c|c|}
\topline
Project code & Frequency & Array & No of sources & Date of observation \\\hline
VLA/15A-068 & 6 GHz & A &  8 & 19th July, 2015\\ 
VLA/16A-144 & 15 GHz & B &  5 & 29th May, 2016\\
VLA/13B-020 & 10 GHz & A &  2 &20th Feb and 14th March, 2014 (archival) \\
\hline
\end{tabular}
\end{table}
\twocolumn
\end{document}